\documentclass[12pt]{iopart}

\expandafter\let\csname equation*\endcsname\relax
\expandafter\let\csname endequation*\endcsname\relax

\usepackage{amsmath, amssymb, graphicx, subfigure, bm, color}
\usepackage[normalem]{ulem}
\usepackage{color, hyperref}
\usepackage[all]{hypcap}

\definecolor{MyDarkGreen}{rgb}{0,0.6,0}
\definecolor{MyDarkBlue}{rgb}{0,0,0.8}
\definecolor{MyDarkRed}{rgb}{0.6,0,0.3}
\hypersetup{breaklinks=true,colorlinks=true,plainpages=true,linktocpage=true,linkcolor=MyDarkBlue,citecolor=MyDarkGreen,urlcolor=MyDarkRed,pdfborder={0 0 0},%
pdfauthor={Sang-Kil Son},%
pdftitle={Determination of multiwavelength anomalous diffraction coefficients at high x-ray intensity}%
}

\newlength{\figurewidth}
\newcommand{\ket}[1]{\ensuremath{\vert#1\rangle}}

\newcommand{\braketoperator}[3]{\ensuremath{\left\langle#1\left\lvert#2\right\rvert#3\right\rangle}}
\newcommand{\braketop}[3]{\ensuremath{\langle#1\lvert#2\rvert#3\rangle}}

\begin{document}

\title[MAD coefficients at high x-ray intensity]{Determination of multiwavelength anomalous diffraction coefficients at high x-ray intensity}

\author{Sang-Kil Son$^{1,2}$, Henry N Chapman$^{1,2,3}$, and Robin Santra$^{1,2,4}$}

\address{$^1$ Center for Free-Electron Laser Science, DESY, Notkestrasse 85, 22607 Hamburg, Germany}
\address{$^2$ The Hamburg Centre for Ultrafast Imaging, Luruper Chaussee 149, 22761 Hamburg, Germany}
\address{$^3$ Department of Physics, University of Hamburg, Luruper Chaussee 149, 22761 Hamburg, Germany}
\address{$^4$ Department of Physics, University of Hamburg, Jungiusstrasse 9, 20355 Hamburg, Germany}
\ead{sangkil.son@cfel.de}
\begin{abstract}
The high-intensity version of multiwavelength anomalous diffraction (MAD) has a potential for solving the phase problem in femtosecond crystallography with x-ray free-electron lasers (XFELs).
For MAD phasing, it is required to calculate or measure the MAD coefficients involved in the key equation, which depend on XFEL pulse parameters.
In the present work, we revisit the generalized Karle-Hendrickson equation to clarify the importance of configurational fluctuations of heavy atoms induced by intense x-ray pulses, and investigate the high-intensity cases of transmission and fluorescence measurements of samples containing heavy atoms.
Based on transmission/fluorescence and diffraction experiments with crystalline samples of known structures, we propose an experimental procedure to determine all MAD coefficients at high x-ray intensity, which can be used in \emph{ab initio} phasing for unknown structures.
\end{abstract}

\submitto{\jpb}
\maketitle

\section{Introduction}\label{sec:introduction}
X-ray free-electron lasers (XFELs)~\cite{Ackermann07,Shintake08,Emma10,Ishikawa12a}, which feature ultraintense and ultrashort x-ray pulses, have brought us a new way of thinking about x-ray--matter interaction and have an impact on various scientific fields, such as atomic and molecular physics~\cite{Young10,Hoener10,Rudek12}, x-ray optics~\cite{Rohringer12}, material science~\cite{Vinko12}, astrophysics~\cite{Bernitt12}, and molecular biology~\cite{Chapman11,Seibert11,Boutet12,Redecke13,Kern13}.
Many collections and reviews on scientific achievements with XFELs are available~\cite{Bucksbaum11,Marangos11,Ullrich12}, including the current special issue on ``Frontiers of FEL Science.''

One of the most prominent XFEL applications is femtosecond x-ray crystallography, which promises to revolutionize structural biology.
The most recent breakthrough in this direction is the first determination of an unknown biological molecular structure with an XFEL~\cite{Redecke13}.
The determination of 3-dimensional macromolecular structures is crucial for understanding their biological functions at the molecular level and for designing new drugs targeting their mechanisms.
However, the key component to reconstruct the molecular structure from an x-ray scattering pattern is the phase of the x-ray scattering amplitude, which is inevitably not measurable in x-ray crystallography experiments.
Note that the molecular replacement technique, which still needs a structurally similar reference structure to phase the new structure, was employed in \cite{Redecke13}.
The phase determination without any previously known structure has been, and still is, a long-lasting challenge in x-ray crystallography~\cite{Karle50,Taylor03}.

The multiwavelength anomalous diffraction (MAD) method~\cite{Guss88,Hendrickson89,Hendrickson91} with synchrotron radiation is one of the major achievements to address this phase problem.
Recently, we proposed a generalized version of MAD phasing at high x-ray intensity~\cite{Son11e}, directly applicable to femtosecond crystallography with an XFEL.
Because of the unprecedentedly high x-ray fluence from an XFEL, individual atoms in a sample undergo multiphoton multiple ionization dynamics, which are characterized by multiple sequences of one-photon ionization accompanied by radiative and/or Auger (Coster-Kronig) decays.
This electronic radiation damage, especially to heavy atoms in a sample, hinders a direct implementation of MAD with an XFEL.
By taking into account the detailed ionization dynamics of heavy atoms during intense x-ray pulses, we demonstrated the existence of a generalized Karle-Hendrickson equation in the high-intensity regime.

Knowing the MAD coefficients involved in this key equation is crucial to determine the phase information.
In \cite{Son11e}, they were calculated using the \textsc{xatom} toolkit~\cite{Son11a} taking into consideration the detailed ionization dynamics of heavy atoms.
These calculated results have convinced us that MAD at high x-ray intensity will work and that dramatic changes in the MAD coefficients at high fluence can be even beneficial for the phase determination.
Currently, the theoretical model of ionization dynamics used in \cite{Son11e} is the only way to determine the MAD coefficients for a given heavy atom.
To test our ability to describe ionization dynamics of heavy atoms embedded in macromolecules, it is necessary to measure the MAD coefficients in experiment and to make quantitative comparisons between theory and experiment.
In this paper, we propose an experimental procedure to determine those MAD coefficients by employing transmission and/or fluorescence and diffraction measurements on known crystalline structures.

The paper is organized as follows.
\Sref{sec:MAD} reviews the generalized Karle-Hendrickson equation.
In \sref{sec:analysis} we analyze the scattering intensity to show the importance of configurational fluctuations induced by intense x-ray pulses.
In \sref{sec:transmission}, we present a derivation of the transmission coefficient in the high-intensity regime.
In \sref{sec:fluorescence}, we discuss x-ray fluorescence yields at high x-ray intensity, as a possible tool to measure one of the MAD coefficients.
In \sref{sec:all_MAD_coeff}, an experimental procedure to determine all MAD coefficients is proposed.
\Sref{sec:conclusion} concludes with a summary.

\section{MAD at high intensity}\label{sec:MAD}
In the conventional MAD phasing method with synchrotron radiation, where electronic damage to heavy atoms is almost negligible, the Karle-Hendrickson equation~\cite{Karle80,Hendrickson85} is the basis for solving the phase problem.
In \cite{Son11e}, we proposed a generalized version of the MAD phasing method including severe electronic damage to heavy atoms, which is applicable at high x-ray intensity.
The key equations are the generalized Karle-Hendrickson equation and its MAD coefficients of $a$, $b$, $c$, and $\tilde{a}$ expressed with population dynamics of electronic configurations of heavy atoms during an x-ray pulse.
In this section, we review the essence of the generalized Karle-Hendrickson equation.
Detailed discussions can be found in \cite{Son11e}.

MAD utilizes the dispersion correction to the elastic x-ray scattering~\cite{Rupp10,Als-Nielsen01}.
Near an inner-shell absorption edge, the atomic form factor depends on the photon energy $\omega$,
\begin{equation}
f(\mathbf{Q},\omega) = f^0(\mathbf{Q}) + f'(\omega) + \rmi f''(\omega),
\end{equation}
where $\mathbf{Q}$ is the photon momentum transfer.
The molecular form factor is given by
\begin{equation}
F^0(\mathbf{Q}) = \sum_{j=1}^N f^0_j(\mathbf{Q}) \rme^{\rmi \mathbf{Q} \cdot \mathbf{R}_j},
\end{equation}
where $N$ is the number of atoms, $f^0_j(\mathbf{Q})$ and $\mathbf{R}_j$ are the normal atomic form factor and the position of the $j$th atom, respectively.
Note that $F^0(\mathbf{Q})$ is a complex number, so it has the amplitude, $|F^0(\mathbf{Q})|$, and the phase, $\phi^0(\mathbf{Q}) = \arg\left[ F^0(\mathbf{Q}) \right]$.
The main task of MAD is to solve $|F^0(\mathbf{Q})|$ and $\phi^0(\mathbf{Q})$ from x-ray scattering patterns.
The scattering intensity (per unit solid angle) is given by
\begin{align}\label{eq:generalized_KH}
\frac{\rmd I(\mathbf{Q},\mathcal{F},\omega)}{\rmd \Omega}
= \mathcal{F} C(\Omega) \Big[
\left| F^0_P(\mathbf{Q}) \right|^2
&
 + 
\left| F^0_H(\mathbf{Q}) \right|^2
\tilde{a}(\mathbf{Q},\mathcal{F},\omega)
\nonumber
\\
& + 
\left| F^0_P(\mathbf{Q}) \right|
\left| F^0_H(\mathbf{Q}) \right|
b(\mathbf{Q},\mathcal{F},\omega)
\cos\left( \phi^0_P(\mathbf{Q}) - \phi^0_H(\mathbf{Q}) \right) 
\nonumber
\\
& + 
\left| F^0_P(\mathbf{Q}) \right| 
\left| F^0_H(\mathbf{Q}) \right| 
c(\mathbf{Q},\mathcal{F},\omega)
\sin\left( \phi^0_P(\mathbf{Q}) - \phi^0_H(\mathbf{Q}) \right) 
\nonumber
\\
& + 
N_H 
\left| f^0_H(\mathbf{Q}) \right|^2
\left\lbrace a(\mathbf{Q},\mathcal{F},\omega) - \tilde{a}(\mathbf{Q},\mathcal{F},\omega) \right\rbrace
\Big],
\end{align}
where $\mathcal{F}$ is the fluence given by the number of photons per unit area, $C(\Omega)$ is a coefficient given by the polarization of the x-ray pulse, and $N_H$ is the number of heavy atoms.
The subscript $P$ refers to light atoms in any protein (or any macromolecule) and the subscript $H$ indicates heavy atoms.
In \eref{eq:generalized_KH}, there are three unknowns to be solved: $|F^0_P(\mathbf{Q})|$, $|F^0_H(\mathbf{Q})|$, and $\phi^0_P(\mathbf{Q}) - \phi^0_H(\mathbf{Q})$ for every $\mathbf{Q}$.
Given the scattering intensity measurements $\rmd I / \rmd \Omega$ at more than three different $\omega$, those unknowns are solved if the MAD coefficients of $a$, $b$, $c$, and $\tilde{a}$ are pre-determined.
These MAD coefficients are given by
\begin{subequations}\label{eq:coefficient}
\begin{align}
\label{eq:a}
a(\mathbf{Q},\mathcal{F},\omega) &= \frac{1}{\left\lbrace f^0_H(\mathbf{Q}) \right\rbrace^2} \sum_{I_H} \bar{P}_{I_H}(\mathcal{F},\omega) \left| f_{I_H}(\mathbf{Q},\omega) \right|^2, 
\\
\label{eq:b}
b(\mathbf{Q},\mathcal{F},\omega) &= \frac{2}{f^0_H(\mathbf{Q})} \sum_{I_H} \bar{P}_{I_H}(\mathcal{F},\omega) \left\lbrace f^0_{I_H}(\mathbf{Q}) + f'_{I_H}(\omega) \right\rbrace,
\\
\label{eq:c}
c(\mathbf{Q},\mathcal{F},\omega) &= \frac{2}{f^0_H(\mathbf{Q})} \sum_{I_H} \bar{P}_{I_H}(\mathcal{F},\omega) f''_{I_H}(\omega),
\\
\label{eq:tilde_a}
\tilde{a}(\mathbf{Q},\mathcal{F},\omega) &= \frac{1}{\left\lbrace f^0_H(\mathbf{Q}) \right\rbrace^2} \int_{-\infty}^{\infty} \!\! \rmd t \, g(t) \left| \sum_{I_H} P_{I_H}(\mathcal{F},\omega,t) f_{I_H}(\mathbf{Q},\omega) \right|^2,
\end{align}
\end{subequations}
where $f_{I_H}$ is the atomic form factor of the $I_H$th electronic configuration of one heavy atom and $g(t)$ is the normalized pulse envelope.
Here $P_{I_H}(\mathcal{F},t)$ is the population of the $I_H$th configuration at time $t$ and given $\mathcal{F}$ and $\omega$, representing electronic radiation damage during an intense x-ray pulse.
$\bar{P}_{I_H}(\mathcal{F},\omega) = \int_{-\infty}^{\infty} \rmd t \, g(t) P_{I_H}(\mathcal{F},\omega,t)$ is the pulse-weighted time-averaged population for the $I_H$th configuration.
The MAD coefficients $a$, $b$, $c$, and $\tilde{a}$ are functions of $\mathcal{F}$ and $\omega$.
Even though $f'_{I_H}$ and $f''_{I_H}$ depend on $\omega$ but not on $\mathcal{F}$, the configuration population dynamics represented by $P_{I_H}$ depend on both $\mathcal{F}$ and $\omega$.
Note that only the difference from \cite{Son11e} is that here we explicitly highlight the $\mathcal{F}$-dependence in all MAD coefficients. 

The key assumptions underlying \eref{eq:generalized_KH} and \eref{eq:coefficient} are:
(a) only heavy atoms scatter anomalously and undergo damage dynamics during an x-ray pulse,
(b) configurational changes occur independently, and
(c) we consider only one species of heavy atoms.
Within those assumptions, \eref{eq:generalized_KH} and \eref{eq:coefficient} were derived after properly averaging over all possible configurations among the $N_H$ heavy atoms.
Assumption (c) can always be fulfilled by choosing suitable materials.
Assumption (b) is reasonably valid when heavy atoms are far apart, because then a change in one atomic site does not affect changes in other heavy atoms.
However, assumption (a) needs to be verified further.
Although the photoabsorption cross section of heavy atoms is usually orders of magnitude larger than that of light atoms, there are much more light atoms than heavy atoms in macromolecules.
In \sref{sec:all_MAD_coeff}, we will come back to this point of how to verify assumption (a).

\section{Analysis of the scattering intensity}\label{sec:analysis}
In this section, we reformulate the scattering intensity by using the \textit{dynamical} form factor and the \textit{effective} form factor of heavy atoms in the sample.
This analysis of the scattering intensity will provide insight on how stochastic changes of the electronic structure of heavy atoms during an intense x-ray pulse affect the scattering intensity.
It will also show that electronic configurational fluctuations of heavy atoms are completely missing if one uses only the \textit{effective} form factor in the expression of the scattering intensity.

In \cite{Son11e}, the \textit{dynamical} form factor of the heavy atom was introduced, which is coherently averaged over $I_H$ at given time $t$,
\begin{equation}
\tilde{f}_{H}(t) = \sum_{I_H} P_{I_H}(t) f_{I_H}.
\end{equation}
Note that the dependencies on $\mathbf{Q}$, $\mathcal{F}$, or $\omega$ are omitted for simplicity. 
By using $\tilde{f}_H(t)$, \eref{eq:generalized_KH} may be written in the form,
\begin{equation}\label{eq:f_tilde}
\frac{\rmd I}{\rmd \Omega} = 
\mathcal{F} C(\Omega) 
\int_{-\infty}^{\infty} \!\!\! \rmd t \, g(t)
\left[
\left| F^0_P + \tilde{f}_{H}(t) \sum_{j=1}^{N_H} \rme^{\rmi \mathbf{Q} \cdot \mathbf{R}_j } \right|^2 
+ N_H V_\text{config}(t)
\right],
\end{equation}
where $V_\text{config}(t)$ is the variance of $\tilde{f}_H$ over different configurations at a given time $t$,
\begin{equation}
V_\text{config}(t) = \sum_{I_H} P_{I_H}(t) | f_{I_H} |^2 - \left| \sum_{I_H} P_{I_H}(t) f_{I_H} \right|^2.
\end{equation}
Then the pulse-weighted time-averaged variance is connected to the last term of \eref{eq:generalized_KH},
\begin{equation}
\bar{V}_\text{config} = \int_{-\infty}^{\infty} \!\!\! \rmd t \, g(t) V_\text{config}(t) = \left( f^0_H \right)^2 \left( a - \tilde{a} \right).
\end{equation}
From \eref{eq:f_tilde} one can easily see that the coherent sum underlies the formation of the Bragg peaks implying that all heavy atoms are described by the same $\tilde{f}_H(t)$ during the time propagation under the x-ray pulse.
On the other hand, the remaining part, $N_H \bar{V}_\text{config}$, represents fluctuations from all different configurations induced by electronic damage dynamics, corresponding to a diffuse background.

Next, we introduce the \emph{effective} form factor of the heavy atom, which is a pulse-weighted time-average of $\tilde{f}_H(t)$,
\begin{equation}
\bar{f}_{H} = \int_{-\infty}^{\infty} \!\!\! \rmd t \, g(t) \tilde{f}_{H}(t) = \sum_{I_H} \bar{P}_{I_H} f_{I_H}.
\end{equation}
Plugging $\bar{f}_H$ into \eref{eq:generalized_KH}, the generalized Karle-Hendrickson equation is rewritten as
\begin{equation}\label{eq:f_bar}
\frac{\rmd I}{\rmd \Omega} = 
\mathcal{F} C(\Omega) 
\left[
\left| F^0_P + \bar{f}_{H} \sum_{j=1}^{N_H} \rme^{\rmi \mathbf{Q} \cdot \mathbf{R}_j } \right|^2 
+ N_H \bar{V}_\text{config} + \left| \sum_{j=1}^{N_H} \rme^{\rmi \mathbf{Q} \cdot \mathbf{R}_j } \right|^2 V_\text{time}
\right],
\end{equation}
where $V_\text{time}$ is the variance of $\tilde{f}_H(t)$ over time,
\begin{equation}
V_\text{time} = \int_{-\infty}^{\infty} \!\!\! \rmd t \, g(t) \left| \tilde{f}_H(t) \right|^2 - \left| \int_{-\infty}^{\infty} \!\!\! \rmd t \, g(t) \tilde{f}_{H}(t) \right|^2.
\end{equation}
In \eref{eq:f_bar}, the first term is calculated using a molecular form factor assuming that all heavy atoms may be described with a single, time-independent scattering factor, $\bar{f}_H$.
The first term in \eref{eq:f_bar}, $\left| F^0_P + \bar{f}_{H} \sum_{j=1}^{N_H} \exp[ \rmi \mathbf{Q} \cdot \mathbf{R}_j ] \right|^2$, would be the simplest expression including electronic radiation damage to heavy atoms.
However, it does not include dynamical fluctuations of configurations during the ionizing x-ray pulse.
Their contributions are proportional not only to $N_H$ via $\bar{V}_\text{config}$ but also to the coherent sum over heavy atoms ($\propto N_H^2$) via $V_\text{time}$.

\begin{figure}
\subfigure[]{\includegraphics[width=6in]{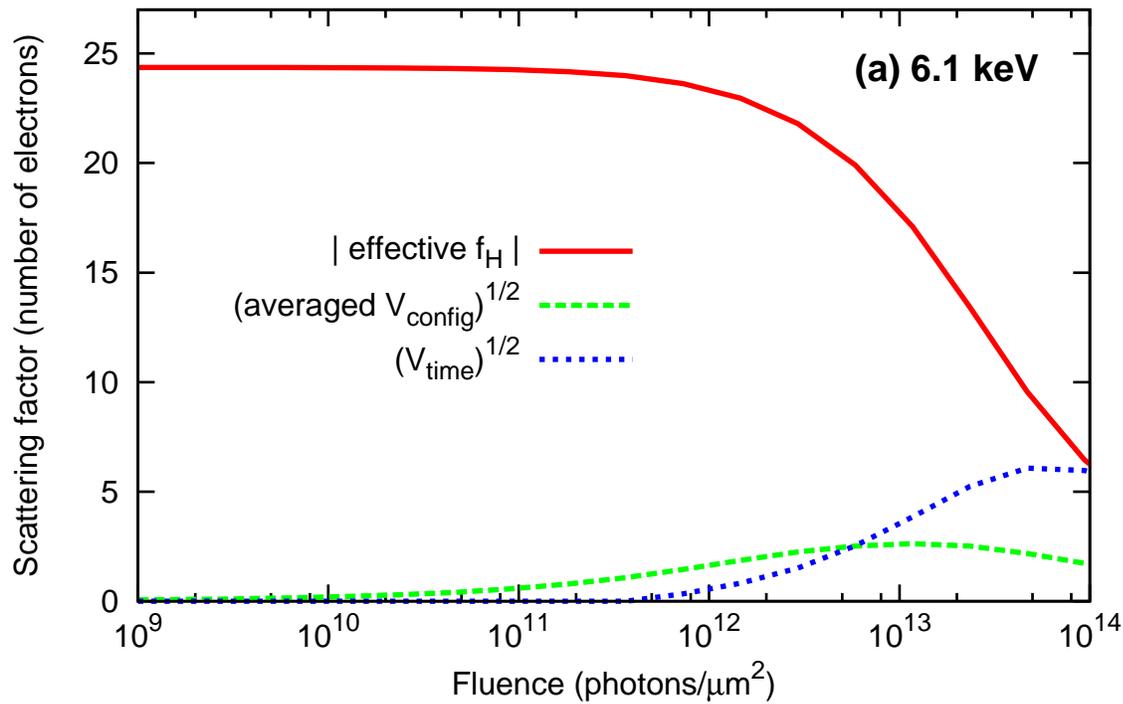}}
\subfigure[]{\includegraphics[width=6in]{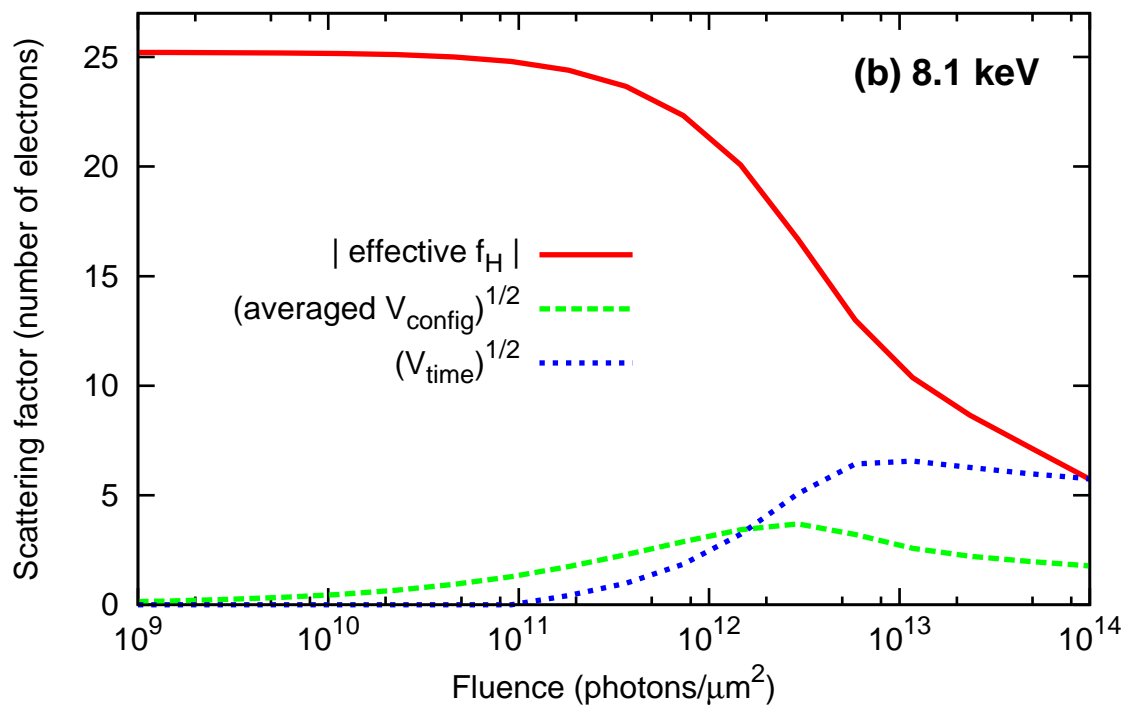}}
\caption{\label{fig:variance}Magnitude of the effective form factor ($| \bar{f}_H |$) and its two different standard deviations ($\bar{V}_\text{config}^{1/2}$ and $V_\text{time}^{1/2}$) of Fe as a function of the fluence, (a) at a photon energy of 6.1~keV (below $K$-edge) and (b) at a photon energy of 8.1~keV (above $K$-edge).
}
\end{figure}

\Fref{fig:variance} shows the magnitude of the effective form factor, $|\bar{f}_H|$, and its two different standard deviations, $\bar{V}_\text{config}^{1/2}$ and $V_\text{time}^{1/2}$, of an iron (Fe) atom at $\mathbf{Q}$=$\mathbf{0}$ as a function of the fluence.
The photon energy is chosen below (6.1~keV) and above (8.1~keV) the $K$-edge of neutral Fe (7.1~keV).
The pulse shape is a flat-top envelope and the pulse duration is 10~fs.
As shown in \fref{fig:variance}, $|\bar{f}_H|$ drops rapidly after about $\mathcal{F}$=$10^{11}$~photons/$\mu$m$^2$.
$\bar{V}_\text{config}^{1/2}$ is relatively small in comparison with $\bar{f}_H$ and it contributes to the diffuse background, which is usually neglected or subtracted out in data analysis.
On the other hand, $V_\text{time}^{1/2}$ becomes considerably large in the high fluence regime.
This fluctuation should not be neglected because it indeed contributes to the Bragg peaks to be measured.
In our calculations, we do not include resonant absorption~\cite{Rudek12}, shakeup and shakeoff processes~\cite{Persson01}, and impact ionization~\cite{Ziaja05,Caleman09}, which would generate further high charge states.
Thus, the dynamical fluctuation effect would be enhanced after these processes are taken into account.
\Fref{fig:variance} demonstrates that, for successful MAD experiments at XFELs, it is necessary to consider detailed analyses of dynamical changes of configurations induced by ionizing x-ray radiation.

\section{Transmission measurement}\label{sec:transmission}
The transmission experiment can directly measure the imaginary part of the scattering factor in the low-intensity x-ray regime.
The $f''$ values of neutral atoms have been measured and tabulated in \cite{Henke93}.
Here we derive the transmission coefficient from a microscopic picture, which will be applicable at high x-ray intensity.
Imposing the same assumptions as used in the high-intensity MAD theory~\cite{Son11e}, we formulate a generalized expression of the transmission coefficient for a thin layer containing heavy atoms exposed to x-rays at high intensity.

First, we calculate the number of photons before and after the interaction of the photons with an atom.
The change in the number of photons, $\Delta N_\text{ph}$, is given by photoabsorption process closely associated with the dispersion correction to elastic x-ray scattering (see the Appendix for detailed derivation),
\begin{equation}\label{eq:Delta_N}
\Delta N_\text{ph} = \frac{4 \pi \alpha}{\omega} J T f''(\omega),
\end{equation}
where $J$ is the photon flux, $T$ is the time interval, and $\alpha$ is the fine-structure constant.
Let us consider a thin layer of $N_A$ atoms, where $N_H$ heavy atoms of the same species are embedded, irradiated by an intense x-ray pulse with a photon energy of $\omega$.
For the intense x-ray pulse, the flux is given by $J(t) = \mathcal{F} g(t)$.
We assume that the sample is thin enough so that all atoms are exposed to the same photon flux.
Individual atoms in the sample are ionized stochastically and their dispersion correction depends on individual electronic configurations at a given time $t$.
Thus, $\Delta N_\text{ph}$ for the intense x-ray pulse is given by
\begin{equation}
\Delta N_\text{ph} = \frac{4 \pi \alpha}{\omega} \int_{-\infty}^{\infty} \! \rmd t \, \mathcal{F} g(t) \sum_I P_I(\mathcal{F},\omega,t) \sum_{j=1}^{N_H} f''_{I_j}(\omega),
\end{equation}
where $j$ denotes the heavy atom index.
In analogy to assumption (a) in \sref{sec:MAD}, we assume that only heavy atoms absorb x-ray photons.
$I$ indicates a global configuration index given by $I = ( I_1, I_2, \cdots, I_{N_H} )$, and $I_j$ indicates the electronic configuration of the $j$th heavy atom.
$P_I(\mathcal{F},\omega,t)$ is the population of the $I$th configuration at time $t$ and given $\mathcal{F}$ and $\omega$.

We also use assumption (b) that the heavy atoms are ionized independently.
Using the same procedure as used to derive the MAD equation~\eref{eq:generalized_KH}, we can further simplify the above expression,
\begin{equation}\label{eq:Delta_N_high_intensity}
\Delta N_\text{ph}
= \frac{4 \pi \alpha}{\omega} \int_{-\infty}^{\infty} \! \rmd t \, \mathcal{F} g(t) N_H \sum_{I_H} P_{I_H}(\mathcal{F},\omega,t) f''_{I_H}(\omega) 
= \frac{4 \pi \alpha}{\omega} \mathcal{F} N_H \sum_{I_H} \bar{P}_{I_H}(\mathcal{F},\omega) f''_{I_H}(\omega).
\end{equation}
Let $\tilde{c}$ be defined by
\begin{equation}
\tilde{c}(\mathcal{F},\omega) = \sum_{I_H} \bar{P}_{I_H}(\mathcal{F},\omega) f''_{I_H}(\omega).
\end{equation}
This is connected to the MAD coefficient $c$ via 
\begin{equation}\label{eq:MAD_c}
c(\mathbf{Q},\mathcal{F},\omega) = \frac{2}{ f^0_H(\mathbf{Q}) } \tilde{c}(\mathcal{F},\omega).
\end{equation}

Given the area $A$ of the thin slab and its infinitesimal thickness $\Delta x$, the fluence is $\mathcal{F} = N_\text{ph} / A$ and the number density of the heavy atoms is $n_H = N_H / ( A \Delta x )$.
Then, \eref{eq:Delta_N_high_intensity} goes over into
\begin{equation}
\Delta N_\text{ph} = \frac{4 \pi \alpha}{\omega} N_\text{ph} n_H \tilde{c}(\mathcal{F},\omega) \Delta x.
\end{equation}
Since $\tilde{c}$ also depends on $N_\text{ph} (=\mathcal{F} A)$, the equation needs to be solved self-consistently.
Thus, if we restrict ourselves to a very thin sample with a finite thickness $x$ such that $\left( 4 \pi \alpha n_H \tilde{c} / \omega \right) x \ll 1$, the x-ray transmission through the sample is approximated by
\begin{equation}\label{eq:transmission}
\frac{N_\text{ph}(x)}{N_\text{ph}(0)} \approx 1 + \frac{4 \pi \alpha}{\omega} n_H \tilde{c}(\mathcal{F},\omega) x.
\end{equation}
Therefore, by measuring the transmission of the thin sample, one can directly obtain the MAD coefficient $c$ at given fluence $\mathcal{F}$ and photon energy $\omega$ in the high-intensity x-ray regime.
The $\mathbf{Q}$-dependence of $c$ is contained in the factor $2 / f^0_H(\mathbf{Q})$.

There are some practical issues in transmission measurement.
It may not be trivial to prepare a thin crystalline sample enough to hold the direct relation between the MAD coefficient $c$ and the transmission coefficient.
For example, if we use a Fe crystalline sample (density: 7.874~g/cm$^3$) with a thickness of 200~nm, the variation of the transmission for a photon energy of 6~keV to 10~keV is estimated as $\sim$6\%.
It is challenging to measure such a small variation when using a high-intensity x-ray beam.

\section{Fluorescence measurement}\label{sec:fluorescence}
In conventional MAD experiments, $f''$ is determined by using the optical theorem~\cite{Als-Nielsen01},
\begin{equation}\label{eq:optical_theorem}
f''(\omega) = -\frac{\omega}{4 \pi \alpha} \sigma_P,
\end{equation}
where $\sigma_P$ is the photoabsorption cross section.
This expression can be verified with a microscopic picture (see the Appendix).
To measure $\sigma_P$, fluorescence measurement has been used~\cite{Rupp10}, because fluorescence signals are proportional to $\sigma_P$ in the low-intensity regime where one-photon absorption is not saturated.
In the case of an intense x-ray pulse generated by an XFEL, however, one-photon absorption may become saturated and non-linear response is expected.
For example, the photoabsorption cross section of neutral Fe at 7.6~keV is $\sim$33~kbarns, so the minimum fluence to saturate one-photon absorption is $\sim$$3.0\times10^{11}$~photons/$\mu$m$^2$.
In the high-intensity regime above this minimum fluence, the fluorescence signal is no more linearly proportional to the number of incident photons~\cite{Rudek12} and not directly connected to the photoabsorption cross section.

\begin{figure}
\includegraphics[width=6in]{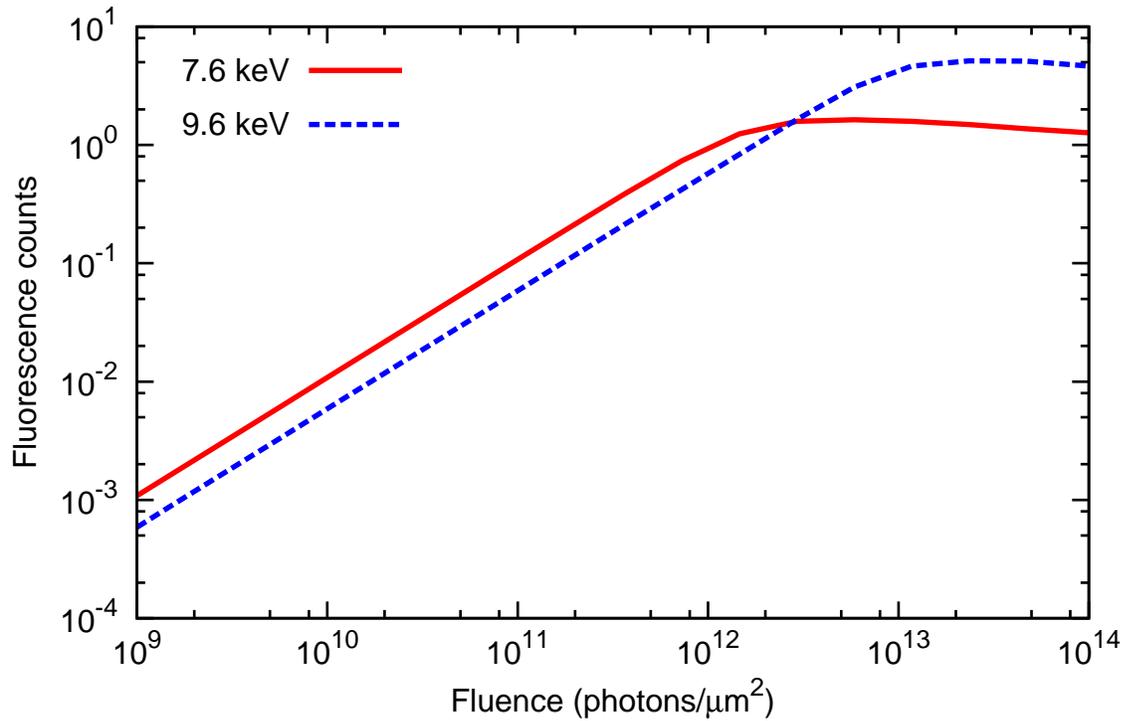}
\caption{\label{fig:fluo_vs_fluence}Number of fluorescence photons per an Fe atom as a function of the fluence.
}
\end{figure}

\Fref{fig:fluo_vs_fluence} shows the number of fluorescence photons, $N_\text{fluo}$, from a single Fe atom as a function of the fluence.
We employ the \textsc{xatom} toolkit to calculate fluorescence counts integrating fluorescence spectra over transition energies~\cite{Son12f,xatom}.
Here, we assume that our sample is thin enough to have no transversal dependence of the photon flux.
To connect fluorescence to $K$-shell absorption of Fe, the fluorescence photons are counted only if the fluorescence energy, $E_\text{fluo}$, is above 6~keV.
The pulse shape used is a flat-top envelope with a temporal width of 10~fs.
The photon energies of 7.6~keV and 9.6~keV are all above the $K$-edge of neutral Fe.
Before saturation, the fluorescence count behaves linearly proportional to the fluence.
After saturation, on the other hand, x-rays keep stripping off electrons from Fe ions and the $K$-shell ionization potential increases as the charge state increases.
Therefore, the $K$-shell ionization is closed earlier when a lower photon energy is used, and the fluorescence count becomes flat and even decreasing if the fluence is too high.

\begin{figure}
\includegraphics[width=6in]{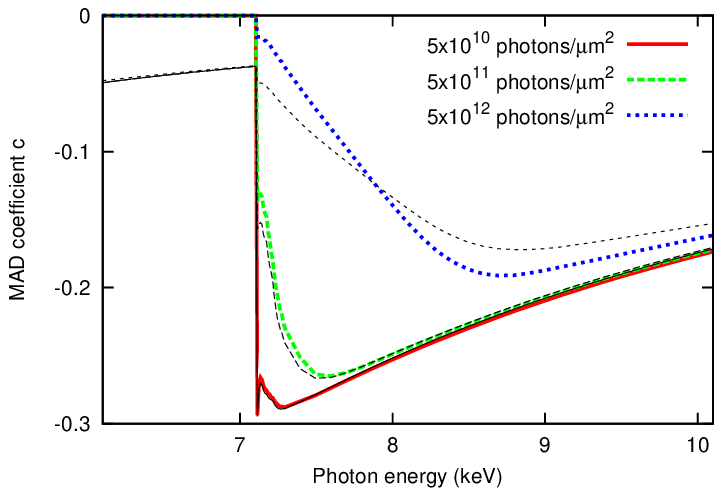}
\caption{\label{fig:fluo_MAD_c}MAD coefficient $c$ obtained from fluorescence yields and calculated by \eref{eq:c} as a function of the incident photon energy at several fluences.
The thick lines are $c$ values converted from the fluorescence yields.
The thin lines are $c$ values calculated by \eref{eq:c}.
}
\end{figure}

\Fref{fig:fluo_MAD_c} compares the MAD coefficient $c$ of Fe, obtained from the fluorescence yield and calculated by \eref{eq:c}, as a function of the photon energy.
The same pulse shape is used as that in \fref{fig:fluo_vs_fluence}.
The fluorescence yields are obtained by the ratio between $N_\text{fluo}$ and the number of incident photons, $N_\text{ph}$.
$N_\text{fluo}$ is counted for $E_\text{fluo} \ge 6$~keV.
Assuming that the fluorescence yield is linearly proportional to $\sigma_P$ and using \eref{eq:optical_theorem}, the MAD coefficient $c$ is converted by
\begin{equation}
\tilde{c}(\mathcal{F},\omega) = \gamma \omega \frac{ N_\text{fluo} }{ N_\text{ph} },
\end{equation}
where $\gamma$ is a single scaling factor applied for all the thick curves in \fref{fig:fluo_MAD_c}, as well as the relation between $c$ and $\tilde{c}$ in \eref{eq:MAD_c}.
In the low-intensity limit, the curves converted from the fluorescence yield (thick) and calculated by \eref{eq:c} (thin) are very similar.
In the high-intensity limit, for example, at the fluence of $5\times10^{12}$~photons/$\mu$m$^2$, the measured $c$ from fluorescence deviates from the calculated $c$ by $\sim$10\% near the peak around 8.6~keV, even though it shows a qualitatively similar trend to the calculated $c$.
\Fref{fig:fluo_MAD_c} demonstrates that it is possible to determine the MAD coefficient $c$ if one scans the photon energy and the fluence when performing the fluorescence measurement.

With a high resolution in fluorescence spectra, one can observe different charge states~\cite{Vinko12} or changes of oxidation states~\cite{Kern13}, which might provide additional information on heavy atoms at high x-ray intensity.
However, the proposed experimental scheme does not require high-resolution fluorescence spectra; instead, it is enough to distinguish emitted photons from light atoms and heavy atoms in order to be able to count fluorescence photons from heavy atoms only.

\section{Determination of the MAD coefficients}\label{sec:all_MAD_coeff}
Once $c$ is determined by use of transmission and/or fluorescence measurement, other MAD coefficients can be obtained as follows.
If one uses a crystal consisting only of the heavy-atom species of interest, then \eref{eq:generalized_KH} is reduced to
\begin{equation}
\frac{\rmd I}{\rmd \Omega}
= \mathcal{F} C(\Omega) \Big[ \left| F^0_H \right|^2 \tilde{a}
 + N_H \left| f^0_H \right|^2 \left( a - \tilde{a} \right) \Big].
\end{equation}
From the Bragg peaks in the diffraction measurement of this sample, one can determine the MAD coefficient $\tilde{a}$ because $\left| F^0_H \right|$ is known.
In this case, we assume that $(a-\tilde{a})$=0.
Then, by carrying out diffraction measurements with simple composite crystals whose structures are already known (for example, FeO, Fe$_2$O$_3$, or Fe$_3$O$_4$), the MAD coefficient $b$ can be determined from \eref{eq:generalized_KH} since all other quantities are known.
Different compositions would give the same $b$ values, if assumption (a) remains valid.
A series of experiments with different compositions would help us understand how neighboring atoms could affect ionization dynamics of the central heavy atom.
The term $(a-\tilde{a})$ would not be easy to quantify in the diffraction experiment, because measurement of the diffuse background with a sufficiently high signal-to-noise ratio is challenging when using crystals.
However, our calculations shown in \fref{fig:variance} guide us how much the $(a-\tilde{a})$ term would contribute to the diffuse background as a function of the fluence and the photon energy.

So far, the high-intensity MAD coefficients have been obtained by numerical simulations of multiphoton multiple ionization dynamics~\cite{Son11e}.
This theoretical description of ionization dynamics has been tested by comparison with recent XFEL experiments for isolated heavy atoms~\cite{Rudek12,Rudek13,Fukuzawa13,Motomura13}.
The model of ionization dynamics agrees well with experimental results when the photon energy is far above the ionization threshold~\cite{Rudek12,Rudek13}.
For MAD experiments, however, it is necessary to use photon energies around the ionization threshold.
On the one hand, the model works very well for isolated atoms.
On the other hand, in a molecular environment, as is the case for MAD, charge rearrangement near highly charged heavy atoms might occur and local plasma formed by emitted electrons might modify ionization dynamics.
To verify the above-mentioned issues, it is important to determine the MAD coefficients under experimental conditions suitable for MAD and to make a quantitative comparison between theory and experiment.
It is worthwhile to emphasize that we are not addressing the validity of the generalized Karle-Hendrickson equation, which always holds in the presence of electronic damage.
Here, we are considering how electronic damage occurs during intense x-ray pulses, which is encoded in the MAD coefficients.
Therefore, experimental determination of the MAD coefficients will test our theoretical model of ionization dynamics causing electronic damage in the sample.
Experimental or theoretical determination of the MAD coefficients for a given set of XFEL pulse parameters plays a key role in the \emph{ab initio} MAD phasing method at high x-ray intensity.

\section{Conclusion}\label{sec:conclusion}
In this paper, we have reanalyzed the generalized Karle-Hendrickson equation, which is the key equation for the multiwavelength anomalous diffraction (MAD) phasing method, emphasizing the importance of configurational fluctuations due to stochastic ionization dynamics of heavy atoms occurring during intense x-ray pulses.
The analysis of fluctuations has assured us that the high-intensity version of the Karle-Hendrickson equation is necessary for phasing in the presence of severe electronic radiation damage.
For successful MAD phasing, it is crucial to obtain the MAD coefficients as a function of x-ray pulse parameters such as the fluence and the photon energy.
We have examined transmission and fluorescence experiments to find a possible way of measuring the MAD coefficients. 
The transmission coefficient at high x-ray intensity, which has been derived with the same assumptions as made for the MAD analysis, provides a direct connection to one of the MAD coefficients.
The fluorescence measurement at high x-ray intensity has been examined for the dependence on the fluence and the photon energy, and we have shown that fluorescence yields can be used to determine one of the MAD coefficients.
We have discussed how to determine all other MAD coefficients in combination with transmission/fluorescence and diffraction measurements.
The MAD method is one of the promising phasing methods for femtosecond x-ray crystallography, which is currently among the most active fields in XFEL science.
This work provides essential steps towards structural determination of macromolecules using XFELs.

\ack
S.K.S.\ thanks Jan Malte Slowik and Thomas White for helpful discussions.

\appendix
\setcounter{section}{1}
\section*{Appendix}
Using nonrelativistic quantum electrodynamics in the same framework as that used for calculating cross sections and decay rates of x-ray-induced processes~\cite{Santra09}, we calculate the number of photons before and after the interaction of the photons with an atom.
The Hamiltonian is written as
\begin{equation}
\hat{H} = \hat{H}_0 + \hat{H}_\text{int},
\end{equation}
where $\hat{H}_0$ describes the unperturbed atomic system and the unperturbed x-ray field.
Employing the principle of minimal coupling and the Coulomb gauge, the interaction Hamiltonian, $\hat{H}_\text{int}$, which describes the interaction between x-ray photon and electron fields, is written as
\begin{align}
\hat{H}_\text{int} 
&= \alpha \int \! \rmd^3 x \, \hat{\psi}^\dagger(\mathbf{x}) \left[ \hat{\mathbf{A}}(\mathbf{x}) \cdot \frac{ \bm{\nabla} }{ \rmi } \right] \hat{\psi}(\mathbf{x}) + \frac{\alpha^2}{2} \int \! \rmd^3 x \, \hat{\psi}^\dagger(\mathbf{x}) \hat{A}^2(\mathbf{x}) \hat{\psi}(\mathbf{x})
\\
&= \hat{H}_{\text{int},1} + \hat{H}_{\text{int},2},
\end{align}
where $\hat{\psi}^\dagger(\mathbf{x})$ $[ \hat{\psi}(\mathbf{x}) ]$ is the electron field creation (annihilation) operator, and $\hat{\mathbf{A}}(\mathbf{x})$ is the vector potential operator.
$\hat{H}_{\text{int},1}$ contains the `$\mathbf{p}\cdot\mathbf{A}$' term and $\hat{H}_{\text{int},2}$ does the `$A^2$' term.
Treating $\hat{H}_\text{int}$ as a perturbation, the state vector in the interaction picture is written as
\begin{equation}
\ket{\Psi, t} = \ket{I} + \ket{\Psi^{(1)}, t} + \ket{\Psi^{(2)}, t} + \cdots,
\end{equation}
where $\ket{\Psi^{(1)}, t}$ and $\ket{\Psi^{(2)}, t}$ are the first- and second-order perturbation corrections, respectively,
\begin{align}
\ket{\Psi^{(1)}, t} 
&= - \rmi \int_{-\infty}^t \! \rmd t' \; \rme^{\rmi \hat{H}_0 t'} \hat{H}_{\text{int},1} \rme^{-\varepsilon |t'|} \rme^{-\rmi E_I t'} \ket{I}
\\
\ket{\Psi^{(2)}, t} 
&= - \rmi \int_{-\infty}^t \! \rmd t' \; \rme^{\rmi \hat{H}_0 t'} \hat{H}_{\text{int},2} \rme^{-\varepsilon |t'|} \rme^{-\rmi E_I t'} \ket{I}
\nonumber
\\
& \quad
- \int_{-\infty}^t \! \rmd t' \; \rme^{\rmi \hat{H}_0 t'} \hat{H}_{\text{int},1} \rme^{-\varepsilon |t'|} \rme^{-\rmi \hat{H}_0 t'} \int_{-\infty}^{t'} \! \rmd t'' \; \rme^{\rmi \hat{H}_0 t''} \hat{H}_{\text{int},1} \rme^{-\varepsilon |t''|} \rme^{-\rmi E_I t''} \ket{I}
\end{align}
The initial state ($t \rightarrow -\infty$) is expressed by
\begin{equation}
\ket{I} = \ket{\Psi_0^{N_\text{el}}} \ket{N_\text{in}},
\end{equation}
where $\ket{\Psi_0^{N_\text{el}}}$ is the electronic ground state with $N_\text{el}$ electrons and $\ket{N_\text{in}}$ is the x-ray photon field with $N_\text{in}$ photons.
For simplicity, we assume that in the incoming state of the photon field, only a single mode is occupied.

Let us define the operator counting the number of photons 
in the incoming mode $(\mathbf{k}_\text{in},\lambda_\text{in})$,
\begin{equation}
\hat{O}_N = \sum_{\lambda_\text{in}} \hat{a}_{\mathbf{k}_\text{in},\lambda_\text{in}}^\dagger \hat{a}_{\mathbf{k}_\text{in},\lambda_\text{in}}^{\phantom\dagger},
\end{equation}
where $\mathbf{k}_\text{in}$ indicates the wave vector in the incoming mode, $\lambda_\text{in}$ denotes its polarization direction, and $\hat{a}_{\mathbf{k},\lambda}^\dagger$ $[ \hat{a}_{\mathbf{k},\lambda} ]$ creates (annihilates) a photon in the mode $(\mathbf{k},\lambda)$.

We calculate the expectation value of $\hat{O}_N$ to obtain the numbers of photons before ($N_\text{in}$) and after ($N_\text{out}$) the interaction of the photons with an atom.
The number of incoming photons is given by
\begin{equation}
\label{eq:N_in}
N_\text{in} = \lim_{t \rightarrow -\infty} \braketoperator{\Psi,t}{\hat{O}_N}{\Psi,t} = \braketoperator{I}{\hat{O}_N}{I}.
\end{equation}
Here, we calculate the expectation value of $\hat{O}_N$ at time $t$ up to the second-order correction,
\begin{equation}
\braketoperator{\Psi,t}{\hat{O}_N}{\Psi,t} = A_{00} + A_{01} + A_{10} + A_{11} + A_{02} + A_{20},
\end{equation}
where $A_{ij} = \braketop{ \Psi^{(i)},t }{ \hat{O}_N }{ \Psi^{(j)}, t }$.
The expansion terms are given by
\begin{align}
A_{00} &= \braketoperator{I}{\hat{O}_N}{I},
\\
A_{01} + A_{10} &= - \rmi \int_{-\infty}^{t} \! \rmd t' \; \rme^{-\varepsilon |t'|} \braketoperator{I}{ \hat{O}_N \hat{H}_{\text{int},1} }{I} + \text{c.c.} = 0,
\\
A_{11} &= \sum_{M \neq I} \left( \int_{-\infty}^t \! \rmd t' \; \rme^{\rmi (E_I - E_M) t' - \varepsilon |t'|} \right) \left( \int_{-\infty}^t \! \rmd t' \; \rme^{- \rmi (E_I - E_M) t' - \varepsilon |t'|} \right) 
\nonumber
\\
& \qquad \qquad \times \left| \braketoperator{M}{ \hat{H}_{\text{int},1} }{I} \right|^2 \braketoperator{M}{ \hat{O}_N }{M}
\nonumber
\\
&= \sum_{M \neq I} \frac{\rme^{2 \varepsilon t}}{\varepsilon^2 + ( E_I - E_M )^2} \left| \braketoperator{M}{ \hat{H}_{\text{int},1} }{I} \right|^2 \braketoperator{M}{ \hat{O}_N }{M},
\\
A_{02} + A_{20} &= 
- \rmi \int_{-\infty}^{t} \! \rmd t' \; \rme^{-\varepsilon |t'|} \braketoperator{I}{ \hat{O}_N \hat{H}_{\text{int},2} }{I} + \text{c.c.}
\nonumber
\\
& \quad - \sum_{M \neq I} \int_{-\infty}^t \! \rmd t' \; \rme^{\rmi (E_I - E_M) t' - \varepsilon |t'|} \int_{-\infty}^{t'} \! \rmd t'' \; \rme^{-\rmi (E_I - E_M) t'' - \varepsilon |t''|} 
\nonumber
\\
& \qquad \qquad \times \braketoperator{I}{ \hat{O}_N \hat{H}_{\text{int},1} }{M} \braketoperator{M}{ \hat{H}_{\text{int},1} }{I} + \text{c.c.}
\nonumber
\\
&= - \sum_{M \neq I} \frac{\rme^{2 \varepsilon t}}{2 \varepsilon \left[ \varepsilon + \rmi ( E_I - E_M ) \right] } \braketoperator{I}{ \hat{O}_N }{I} \braketoperator{I}{ \hat{H}_{\text{int},1} }{M} \braketoperator{M}{ \hat{H}_{\text{int},1} }{I} + \text{c.c.}
\nonumber
\\
&= - \sum_{M \neq I} \frac{\rme^{2 \varepsilon t}}{\varepsilon^2 + ( E_I - E_M )^2} \left| \braketoperator{M}{ \hat{H}_{\text{int},1} }{I} \right|^2 \braketoperator{I}{ \hat{O}_N }{I}.
\end{align}
Thus, the number of photons at time $t$ is given by
\begin{align}
\braketoperator{\Psi,t}{\hat{O}_N}{\Psi,t} = \braketoperator{I}{ \hat{O}_N }{I} + \sum_{M \neq I} \frac{e^{2 \varepsilon t}}{\varepsilon^2 + ( E_I - E_M )^2} \left| \braketoperator{M}{ \hat{H}_{\text{int},1} }{I} \right|^2
\nonumber
\\
\times \left[ \braketoperator{M}{ \hat{O}_N }{M} - \braketoperator{I}{ \hat{O}_N }{I} \right].
\end{align}
After adiabatic switching, the number of photons detected is given by
\begin{align}
N_\text{out} 
= \lim_{\substack{t \rightarrow \infty \\ \varepsilon \rightarrow 0^+} } \braketoperator{\Psi,t}{\hat{O}_N}{\Psi,t} 
= \braketoperator{I}{ \hat{O}_N }{I} + \sum_{M \neq I} 2 \pi T \delta( E_I - E_M ) \left| \braketoperator{M}{ \hat{H}_{\text{int},1} }{I} \right|^2 
\nonumber
\\
\times \left[ \braketoperator{M}{ \hat{O}_N }{M} - \braketoperator{I}{ \hat{O}_N }{I} \right],
\end{align}
where $T$ is the time interval.
When $\hat{H}_{\text{int},1}$ with $\ket{M} = \ket{\Psi_M^{N_\text{el}}} \ket{N_\text{in}-1}$ contributes to $N_\text{out}$, the $A_{11}$ term is related to the photoabsorption [$\braketop{M}{ \hat{O}_N }{M} = N_\text{in} - 1$] and the $A_{02} + A_{20}$ terms are related to the dispersion correction [$\braketop{I}{ \hat{O}_N }{I} = N_\text{in}$].
The contribution from the creation of a photon is not allowed because the initial electronic state is given by the ground state.

The photoabsorption cross section $\sigma_P$ and the imaginary part of the scattering factor $f''$ are calculated by
\begin{align}
\sigma_P &= \frac{1}{J} \sum_{M \neq I} 2 \pi \delta( E_I - E_M ) \left| \braketoperator{M}{ \hat{H}_{\text{int},1} }{I} \right|^2,
\\
f''(\omega) &=  - \frac{\omega}{4 \pi \alpha J} \sum_{M \neq I} 2 \pi \delta( E_I - E_M ) \left| \braketoperator{M}{ \hat{H}_{\text{int},1} }{I} \right|^2,
\end{align}
where $\ket{M} = \ket{\Psi_M^{N_\text{el}}} \ket{N_\text{in}-1}$ and $J$ is the x-ray photon flux.
Thus, they are related to each other via
\begin{equation}
f''(\omega) = -\frac{\omega}{4 \pi \alpha} \sigma_P.
\end{equation}
Consequently, the change in the number of photons per atom is given by
\begin{equation}
\Delta N_\text{ph} = N_\text{out} - N_\text{in} = - J T \sigma_P,
\end{equation}
or equivalently,
\begin{equation}
\Delta N_\text{ph} = \frac{4 \pi \alpha}{\omega} J T f''(\omega),
\end{equation}
which proves \eref{eq:Delta_N} in the main text.
For a single atom, $J T \sigma_P$ corresponds to the probability of absorbing one photon.

\section*{References}

\providecommand{\newblock}{}

\end{document}